\documentclass[showpacs,twocolumn]{revtex4}

\usepackage{amsfonts}
\usepackage{graphicx}

\begin{document}

\title{A mean-field kinetic lattice gas model of electrochemical cells}

\author{Marc-Olivier Bernard}
\author{Mathis Plapp}
\author{Jean-Fran\c cois Gouyet}

\affiliation{Laboratoire de Physique de la Mati\`{e}re Condens\'{e}e,\\
CNRS/Ecole Polytechnique, 91128 Palaiseau, France}

\date{\today}

\begin{abstract}
We develop Electrochemical Mean-Field Kinetic Equations (EMFKE) to 
simulate electrochemical cells. We start from a microscopic
lattice-gas model with charged particles, and build mean-field kinetic
equations following the lines of earlier work for neutral particles. We
include the Poisson equation to account for the influence of the electric
field on ion migration, and oxido-reduction processes on the electrode
surfaces to allow for growth and dissolution. We confirm the viability of
our approach by simulating (i) the electrochemical equilibrium at flat
electrodes, which displays the correct charged double-layer, (ii) the growth
kinetics of one-dimensional electrochemical cells during growth and
dissolution, and (iii) electrochemical dendrites in two dimensions.
\end{abstract}


\maketitle

\section{Introduction}

Electrochemical phenomena are ubiquitous in nature and technology. They play
a fundamental role in many materials science problems of high practical
relevance, such as corrosion, electrodeposition of parts and circuitry, and
battery technology \cite{BattCorrWeb}. Electron transfer is also crucially
involved in many biochemical reactions \cite{BioElec}. For physicists,
electrodeposition is interesting because it can lead to the spontaneous
formation of highly complex self-organized patterns. Such branched
aggregates that are often found on the surfaces of natural minerals have
fascinated scientists for centuries because of their plant-like appearance 
\cite{Fleury}.

\begin{figure}[ht]
\begin{center}
\includegraphics[width=8cm]{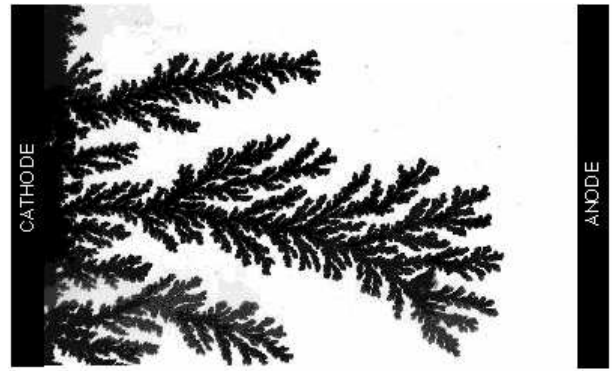}
\end{center}
\caption{Electrodeposition of copper on a glass substrate (reprinted with
permission from V.\ Fleury). The quasi-two-dimensional cell is made of
two copper electrodes and a CuSO$_{4}$ solution (sample size 
$3\times 2\,\mathrm{mm}^{2}$). Only a small part of the sample is shown. }
\label{figfleury}
\end{figure}

\begin{figure}[ht]
\begin{center}
\includegraphics[width=8cm]{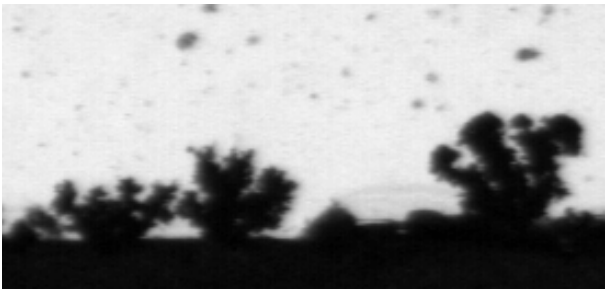}
\end{center}
\caption{Dendritic growth at the interface between a lithium electrode and a
polymer electrolyte (Claire Brissot, PhD thesis, Ecole Polytechnique,
Palaiseau, France, 1998)}
\label{figrosso}
\end{figure}

Under well-controlled laboratory conditions, a variety of different patterns
can be generated, ranging from compact crystals to highly branched dendritic
aggregates that can be fractals (see Figure \ref{figfleury}) or densely
branched. While dendrites rarely appear in traditional applications of
electrodeposition such as metalization and electroplating that take place
close to equilibrium, they can be of practical importance far from
equilibrium, for example in battery technology \cite{Selim74}: Figure \ref
{figrosso} shows the growth of dendrites in a lithium battery. Since
dendrites can perforate insulating layers and lead to shortcuts when they
reach the counterelectrode, they must be eliminated as much as possible. To
achieve this goal, a good understanding of the various phenomena occurring
in electrochemical cells is necessary. The purpose of our paper is to
develop a microscopic model for electrodeposition that can be used to 
elucidate some of the aspects of dendrite formation.

To fix the ideas, let us consider the simplest possible situation that
corresponds to the experimental setup of Figs.~\ref{figfleury} and~\ref
{figrosso}: two electrodes made of the same metal are plunged in an
electrolyte containing ions of the same metal. When a potential difference
is applied by an external generator, as a response an ionic current flows
through the electrolyte. Close to the cathode, ions are reduced by electron
transfer from the electrode and form a growing deposit. The inverse takes
place at the anode: metal is dissolved, and new ions are formed.
For liquid electrolytes (such as used in Fig.~\ref{figfleury}),
the inhomogeneities in the ion concentrations lead to strong convective
flows \cite{Laitinen39,Barkey94,Huth95,Rosso99}. In contrast, for gel-like
electrolytes (such as used in Fig.~\ref{figrosso}), convection is suppressed 
\cite{Rosso94}, and the important ingredients for the description of the ion 
transport in the electrolyte are diffusion and migration of the ions. At the
metal-electrolyte interface, there exists in general a charged double layer
that is much thicker than the microscopic solid-electrolyte interface, but
much smaller than a typical cell dimension. There is a considerable amount
of work on electrochemical cells based on a macroscopic viewpoint \cite{Schm}
in which the microscopic interface is replaced by a mathematically sharp
surface. Such models are generally unable to handle the geometrical
complexity of a fully developed dendrite. In addition, this approach has to
rely on phenomenological models to incorporate reaction kinetics at the
interfaces.

In the statistical physics community, considerable interest in
electrodeposition was spurned by the fact that the geometrical structure of
fractal electrodeposits \cite{Matsu84,Grier86,Sawada86,FleuryChaza} is
strikingly similar to patterns generated by the diffusion-limited
aggregation (DLA) model \cite{DLA}, in which random walkers irreversibly
stick to the growing aggregate. Subsequently, many refinements of the DLA
model were developed to generate various patterns, for example by the
inclusion of surface tension \cite{Vicsek84}, anisotropic growth rules and
noise reduction \cite{Nittmann86,Eckmann89}, introduction of a uniform drift
to mimic an electric field \cite{Meakin83,Jullien84,Hill97}, and
combinations of discrete aggregation and continuous convection models to
study the influence of fluid convection \cite{Marshall}. However, these
models greatly simplify the underlying physics (for example, the charged
double layers are not included) and can hence not yield detailed information
on the relation between growth conditions and characteristics of the growth
structures such as growth speed, branch thickness, and overall structure.
The same is true for a recent mean-field model that does not explicitly
contain the charged ions \cite{Elez}.

Our goal is to build a simplified microscopic model for electrodeposition
that includes enough of the salient physics to make contact with the
macroscopic view of electrochemistry. The electrochemical mean-field kinetic
equations (EMFKE) developed here are based on a lattice-gas model with
simple microscopic evolution rules that contains charged particles, coupled
to a discretized version of the Poisson equation. Lattice gas models have
been used previously in the context of electrochemistry to simulate phenomena 
located on the electrode surfaces, such as adsorption or underpotential 
deposition \cite{rikvold0},
and for studies of ionic transport at liquid-liquid interfaces \cite{Schm2};
however, there exists, to our knowledge, no theoretical study of the
behavior of an entire electrochemical cell based on a microscopic model.

To investigate the dynamics of the lattice model, we extend the formalism of
Mean-Field-Kinetic-Equations (MFKE) \cite{Martin90,Gouyet93,Vaks94} that has
been used to study numerous transport and growth phenomena in alloys,
including diffusion and ordering kinetics \cite{Gouyet95}, spinodal
decomposition \cite{Dobr96,PlappPRL} and dendritic growth \cite{Plapp}; some
preliminary results on the extension to electrochemistry have been published
in Refs.~\cite{Bernard1,Bernard2}. The key feature of this approach 
is that the microscopic particle currents can be written in the 
mean-field approximation as the product of a mobility times the 
gradient of chemical potentials, the latter being the appropriate 
thermodynamic driving forces. The formalism can be generalized
in a natural way to charged particles. The driving forces for particle 
currents are then the gradients of the electrochemical potential.
As a consequence, the resulting model displays the correct 
electrochemical equilibrium at the interfaces. While it
obviously still contains strong simplifications, it is much closer to the
basic microscopic physics and chemistry than the DLA-type models and allows
us to establish a direct link between a microscopic model and the
well-established macroscopic phenomenological equations. Quite remarkably,
the equations of motion of our model also share many common features with a
phase-field formulation of the electrochemical interface that was developed
very recently \cite{GBW}. This indicates that mean-field equations of
simplified models can also be useful to understand the connection between
phase-field models and a more microscopic viewpoint.

In this paper, we will first present the lattice gas model and derive the
electrochemical mean-field kinetic equations (Section 2). In section 3, we 
carry out one-dimensional simulations to demonstrate that the model leads 
to the correct equilibrium at the electrode-electrolyte interface. We
also calculate one-dimensional steady state solutions for moving interfaces
that are in good agreement with the macroscopic continuous model of
Chazalviel \cite{Chaza90}. Finally, we show some preliminary simulations
of dendritic growth in two-dimensional cells. Section 4 contains
a brief discussion and conclusion.

\section{Model}

\subsection{Lattice gas model}

We consider an electrochemical cell made of a dilute binary electrolyte and
two metallic electrodes of the same metal. No supporting electrolyte is
included in the present study. These conditions correspond to the 
experimental situation in Figs. \ref{figfleury} and \ref{figrosso}
(more precisely, to Fig.~\ref{figrosso} since we do not include
convection in our model). The electrodes are modeled by a lattice
that reflects the underlying crystalline structure, and whose sites are
occupied by metallic atoms or vacancies. It is convenient to represent the
electrolyte by the same lattice, occupied by a solvent, cations, anions, or
vacancies. Although there is no physical lattice in the liquid, its presence
here does not play a role due to the high dilution of the ions.

\begin{figure}[tbp]
\begin{center}
\includegraphics[width=8cm]{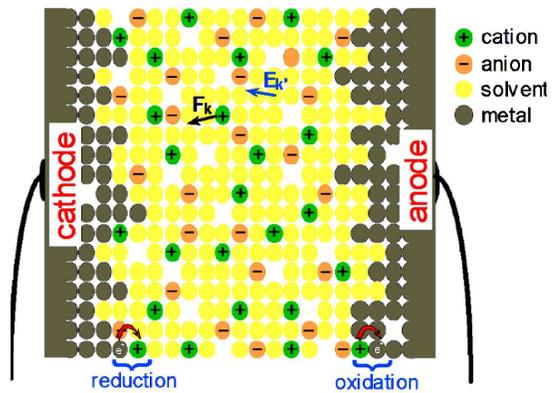}
\end{center}
\caption{The lattice gas model. A fixed potential difference is applied
across the cell. The ions in the electrolyte are submitted to an electric
field $\mathbf{E}_{\mathbf{k}}$ (and hence a force $\mathbf{F}_{\mathbf{k}}=q%
\mathbf{E}_{\mathbf{k}}$) at their lattice site position $\mathbf{k}$. The
various species have short range interactions (here, attractive interactions
are considered between solvent and ions, solvent and solvent, and metal and
metal). Electron transfer takes place on the electrode surfaces.}
\label{figcell}
\end{figure}

For simplicity, we consider here a two-dimensional lattice gas on a square
lattice with lattice spacing $a$ (see Figure \ref{figcell}). The cations 
$M^{+}$ give metallic atoms $M^{0}$ after reduction, while the anions $A^{-}$
are supposed to be non electroactive. The solvent $S$ is neutral, but can
interact through short-range interactions with the other species and with
itself. We specify a microscopic configuration by the set $\{n\}$ of the
occupation numbers $n_{\mathbf{k}}^{\alpha }$ on each site $\mathbf{k}$: $n_{%
\mathbf{k}}^{\alpha }=1$ if $\mathbf{k}$ is occupied by species $%
\alpha=M^{0},M^{+},A^{-},S$ or a vacancy $v$, and $0$ otherwise. We suppose
steric exclusion between the different species, that is, a given site can be
occupied by only one species or it can be empty (vacancy): 
\begin{equation}
\sum_{\alpha }n_{\mathbf{k}}^{\alpha }+n_{\mathbf{k}}^{v}=1 \,.  \label{eq2}
\end{equation}
Finally, as will become clear below, it is convenient to introduce electrons
in the metallic electrodes. They have a particular status that will be
discussed later.

Two very different types of interactions have to be considered. The
short-range interactions (including, for example, van der Waals 
forces, solvatation effects, and chemical interactions) are modeled 
here by nearest-neighbor interactions $\epsilon^{\alpha\beta}$ between
species $\alpha$ and $\beta$ (with the convention that a positive
$\epsilon^{\alpha \beta }$ corresponds to an attractive interaction). 
Interaction energies with vacancies are taken to 
be zero. To take into account the long-range electrostatic 
interaction, it would be possible to introduce appropriate
interaction energies with farther neighbors; however, this procedure 
becomes cumbersome with increasing interaction range. Instead, we
consider a ``coarse-grained'' electrical potential $V_{\mathbf{k}}$ defined on
the lattice sites $\mathbf{k}$ (that is, a potential that has been smoothed
over distances smaller than $a$) and that obeys Poisson's equation with the
simplest nearest-neighbor discretization for the Laplace operator, 
\begin{equation}
\sum_{\mathbf{a}}V_{\mathbf{k+a}}-4V_{\mathbf{k}}=-\frac{a^{2-d}}{\epsilon }%
\sum_{\alpha =+,-}q^{\alpha }n_{\mathbf{k}}^{\alpha }  \label{eqpoisson0}
\end{equation}
where $\epsilon$ is the dielectric constant (for simplicity, we take a
constant that is the same for all species), $q^\alpha$ is the electric
charge of species $\alpha$, and $d$ the spatial dimension. This equation has
to be solved in the electrolyte, that is, outside of the metal clusters
connected to the ends of the cell, and subject to the boundary condition of
constant potential in each electrode.

\begin{figure}[tbp]
\begin{center}
\includegraphics[width=8cm]{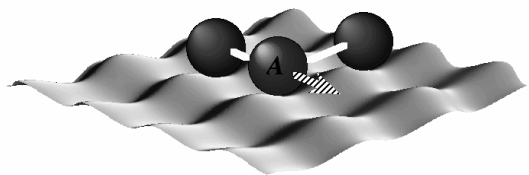}
\end{center}
\caption{In the lattice gas model, an A atom makes activated jumps to empty
nearest neighbor sites. The barrier it has to overcome depends on its
interactions (white links) with its nearest neighbor atoms.}
\label{figbonds}
\end{figure}

\begin{figure}[tbp]
\begin{center}
\includegraphics[width=8cm]{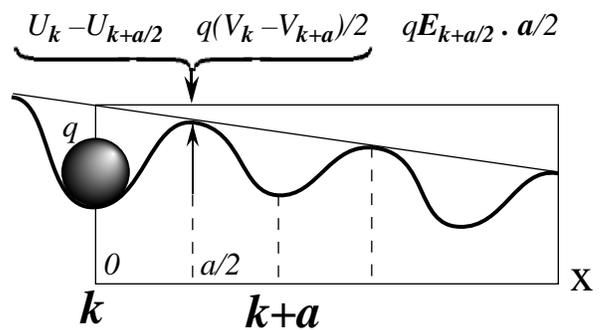}
\end{center}
\caption{In presence of an electric field $\mathbf{E}$, a potential energy
which varies, to first order, like $q\mathbf{E}.\mathbf{x}$ along the jump
path $\mathbf{x}$, is superimposed to the local potential seen by a moving
particle (with charge $q$).}
\label{figfield}
\end{figure}

The configurations evolve when one particle (metal, ion, or solvent) jumps
to one of its vacant nearest neighbor sites. In principle, we could also
include an exchange process between occupied nearest neighbor sites.
This leads to more complicated kinetic equations and has not been
considered here. To specify the jump rates,
we assume that the atoms perform activated jumps. The height of the
activation barrier depends on the local binding energy, that is, the number
and type of bonds that need to be broken (see Fig. \ref{figbonds}), and, for
charged particles, on the local electric field that shifts the barrier
height as shown in Fig. \ref{figfield}. The result for the jump rate from
site $\mathbf{k}$ to site $\mathbf{k+a}$ is 
\begin{eqnarray}
\widetilde{w}_{\mathbf{k,k+a}}^{\alpha }(\{n\}) & = & w^\alpha \exp \left(
-\frac{1}{kT} \sum_{\beta }\sum_{\mathbf{a^{\prime}}}\epsilon ^{\alpha \beta
}n_{\mathbf{k+a^{\prime}}}^{\beta }\right)\nonumber\\
 & & \quad\mbox{}\times \exp \left( \frac{q^{\alpha }}{2kT}
(V_{\mathbf{\ k}}-V_{\mathbf{k+a^{\prime}}})\right) ,  \label{eqjumprates}
\end{eqnarray}
where $kT$ is the (fixed) thermal energy, and $w^\alpha$ is a fixed 
jump frequency that may be different for each species. Here and in the 
following, symbols with tildes will denote electrochemical quantities 
that are sensitive to the electric potential.

\subsection{Mean-field kinetic equations}

The derivation of the Electrochemical Mean-Field Kinetic Equations
\mbox{(EMFKE)} follows the same procedure as for neutral particles \cite
{Martin90,Gouyet93,Vaks94}. We first write the Boolean kinetic equations on
the lattice, starting from the general master equation, 
\begin{eqnarray}
\frac{\partial }{\partial t}P(\{n\},t) & = & \sum_{\{n^{\prime }\}}\left[
W(\{n^{\prime }\}\rightarrow \{n\})P(\{n^{\prime }\},t)\right. \nonumber \\
&&\quad\left.\mbox{}-W(\{n\}\rightarrow\{n^{\prime }\})P(\{n\},t)\right]  
\label{eq1}
\end{eqnarray}
which gives the probability $P$ to find a given configuration $\{n\}$ at
time $t$. $W(\{n\}\rightarrow \{n^{\prime }\})$ is the rate of evolution
from configuration $\{n\}$ to configuration $\{n^{\prime }\}$. We are
interested in the time evolution of the average concentrations of the
various species ($\alpha =+,-,0,S,v)$, 
\begin{equation}
p_{\mathbf{k}}^{\alpha }(t)=\left\langle n_{\mathbf{k}}^{\alpha
}\right\rangle_t =\sum_{\{n\}}n_{\mathbf{k}}^{\alpha }P(\{n\},t).
\label{eq3}
\end{equation}
In the absence of electrochemical processes at the electrodes, the number of
each type of particles remains constant, and the kinetic equation of the
average concentration has the structure of a conservation equation, 
\begin{equation}
\frac{\partial p_{\mathbf{k}}^{\alpha }}{\partial t}=-\sum_{\mathbf{a}}
\widetilde{J}_{\mathbf{k,k+a}}^{\alpha },  \label{eq3a}
\end{equation}
with the current of species $\alpha$ on link $\{\mathbf{k,k+a\}}$ defined
by, 
\begin{equation}
\widetilde{J}_{\mathbf{k,k+a}}^{\alpha }=
\left\langle \widetilde{w}_{\mathbf{k,k+a}}^{\alpha}
  (\{n\})n_{\mathbf{k}}^{\alpha }n_{\mathbf{k+a}}^{v}-
\widetilde{w}_{\mathbf{k+a,k}}^{\alpha }(\{n\})n_{\mathbf{k+a}}^{\alpha}
n_{\mathbf{k}}^{v}\right\rangle,  
\label{eq3b}
\end{equation}
with $\widetilde{w}_{\mathbf{k,k+a}}^{\alpha }(\{n\})$ given by Eq. (\ref
{eqjumprates}). The factor $n_{\mathbf{k}}^{\alpha }n_{\mathbf{k+a}}^{v}$
(and $n_{\mathbf{k+a}}^{\alpha }n_{\mathbf{k}}^{v}$ for the reverse jump)
means that for a jump of $\alpha$ to be possible, the start site must be
occupied by species $\alpha$, while the target site must be empty (occupied
by a vacancy $v$).

In the mean-field approximation, the occupation numbers $n_{\mathbf{k}%
}^{\alpha }$ in the above expressions are replaced by their average $p_{%
\mathbf{k}}^{\alpha }$. This replacement is not unique because of different
possible choices for the factorization of the occupation number operators 
\cite{Gouyet93}. A convenient choice \cite{Gouyet93,Plapp} is the direct
replacement of all occupation numbers by their averages in Eqs. (\ref{eq3b})
and (\ref{eqjumprates}), which leads to 
\begin{eqnarray}
\widetilde{J}_{\mathbf{k,k+a}}^{\alpha} & = & w^\alpha 
\Biggl[ p_{\mathbf{k}}^{\alpha }p_{\mathbf{k+a}}^{v} \exp \Bigr( -\frac{1}{kT}
\sum_{\beta }\sum_{\mathbf{a^{\prime}}} \epsilon ^{\alpha \beta }p_{\mathbf{%
k+a^{\prime}}}^{\beta } \nonumber \\
& & \qquad\qquad\qquad  \mbox{} 
  + \frac{q^{\alpha }}{2kT}(V_{\mathbf{k}}-V_{\mathbf{k+a}})\Bigr)\nonumber \\
& & \mbox{} - p_{\mathbf{k+a}}^{\alpha }p_{\mathbf{k}}^{v} \exp \Bigl( -\frac{%
1}{kT} \sum_{\beta }\sum_{\mathbf{a^{\prime}}} \epsilon ^{\alpha \beta }p_{%
\mathbf{k+a+a^{\prime}}}^{\beta } \nonumber \\
& & \qquad\qquad\qquad \mbox{}
  + \frac{q^{\alpha }}{2kT}(V_{\mathbf{k+a}}-V_{\mathbf{k}})\Bigr) \Biggr].
\end{eqnarray}
This expression can be rewritten without any further approximation as the
product of an electrochemical bond mobility 
$\widetilde{M}_{\mathbf{ij}}^{\alpha }$ times the (discrete) gradient 
of an electrochemical potential $\widetilde{\mu }_{\mathbf{i}}^{\alpha }$, 
\begin{equation}
\widetilde{J}_{\mathbf{k,k+a}}^{\alpha }=-\widetilde{M}_{\mathbf{k,k+a}
}^{\alpha }\mathfrak{D}_{\mathbf{a}}\widetilde{\mu }_{\mathbf{k}}^{\alpha },
\label{eq10}
\end{equation}
where $\mathfrak{D}_{\mathbf{a}}$ is a \textit{difference operator} acting on
the site coordinates, $\mathfrak{D}_{\mathbf{a}}f_{\mathbf{k}}=f_{\mathbf{k+a}%
}-f_{\mathbf{k}}$. The electrochemical potential, 
\begin{eqnarray}
\widetilde{\mu }_{\mathbf{k}}^{\alpha } & = & \mu _{\mathbf{k}}^{\alpha
}+q^{\alpha }V_{\mathbf{k}}\nonumber \\
 & = & -\sum_{\beta }\sum_{\mathbf{a}}\varepsilon
^{\alpha \beta }p_{\mathbf{k+a}}^{\beta }+kT\ln \left( \frac{p_{\mathbf{k}%
}^{\alpha }}{p_{\mathbf{k}}^{v}}\right) +q^{\alpha }V_{\mathbf{k}},
\label{eq11}
\end{eqnarray}
is the sum of three contributions, a local energy due to the interaction of
species $\alpha$ with its local environment, an entropy term (these two
constitute the chemical potential $\mu _{\mathbf{k}}^{\alpha }$), and an
electrostatic energy $q^{\alpha }V_{\mathbf{k}}$. The presence of the
vacancy concentration in the denominator of the entropic contributions comes
from the constraint of Eq.~(\ref{eq2}). The mobility along a bond $\mathbf{i}
$ -$\mathbf{j}$ is given by 
\begin{equation}
\widetilde{M}_{\mathbf{k,k+a}}^{\alpha }=\frac{w^{\alpha }}{kT}p_{\mathbf{k}%
}^{v}p_{\mathbf{k+a}}^{v}\exp \frac{(\widetilde{\mu }_{\mathbf{k}}^{\alpha }+
\widetilde{\mu }_{\mathbf{k+a}}^{\alpha })}{2kT}\mathrm{shc}
\frac{\mathfrak{D}_{\mathbf{a}}\widetilde{\mu }_{\mathbf{k}}^{\alpha }}{2kT},
\label{eq13}
\end{equation}
where we have used the notation $\mathrm{shc~}u=\sinh u/u$ (close to
equilibrium, $\widetilde{\mu }_{\mathbf{k+a}}^{\alpha }\cong \widetilde{\mu }%
_{\mathbf{k}}^{\alpha }$ and $\mathrm{shc}\left[ \mathfrak{D}_{\mathbf{a}}%
\widetilde{\mu }_{\mathbf{k}}^{\alpha }/2kT\right] \cong 1$ ).

In a dilute electrolyte, where the concentration of species $\alpha =+,-,0,$ 
and $v$ is low, neglecting all the terms of order larger than two in the 
concentrations and discrete gradients, the current 
$\widetilde{J}_{\mathbf{k,k+a}}^{\alpha }$ simplifies, 
\begin{equation}
\widetilde{J}_{\mathbf{k,k+a}}^{\alpha }=-w_{\mathbf{k}}^{\alpha }\left( 
\mathfrak{D}_{\mathbf{a}}p_{\mathbf{k}}^{\alpha }+
\frac{p_{\mathbf{k}}^{\alpha }}{p_{\mathbf{k}}^{v}}
\mathfrak{D}_{\mathbf{a}}p_{\mathbf{k}}^{v}+\frac{q^{\alpha
}}{kT}p_{\mathbf{k}}^{\alpha }\mathfrak{D}_{\mathbf{a}}V_{\mathbf{k}}\right) ,
\label{eq10a}
\end{equation}
where $w_{\mathbf{k}}^{\alpha }=w_{0}^{\alpha }\exp \left( -(1/kT)
\sum_{\beta }\sum_{\mathbf{a^{\prime }}}\epsilon ^{\alpha \beta }p_{\mathbf{%
\ k+a^{\prime }}}^{\beta }\right)$. This is the discrete form of the
continuous macroscopic expression
\begin{equation}
\mathbf{j}^{\alpha }=-D^{\alpha \alpha }\mathbf{grad~}c^{\alpha }-D^{\alpha
v}\mathbf{grad~}c^{v}+D^{\alpha \alpha }\frac{q}{kT}c\mathbf{E},
\label{diffmig}
\end{equation}
with a concentration-dependent diffusion coefficient 
$D^{\alpha \alpha }=a^{2}w^{\alpha}$, an off-diagonal diffusion 
coefficient $D^{\alpha v}=a^{2}w^{\alpha }c^{\alpha }/c^{v}$ 
associated with the gradient of the vacancy concentration, and the 
correspondence between a $d$-dimensional cubic lattice and the 
$d-$dimensional continuous space $J_{\mathbf{k,k+a}}\Rightarrow
a^{d-1}\mathbf{j},p_{\mathbf{k}}\Rightarrow a^{d}c$. When the
vacancy concentration is homogeneous, the off-diagonal term is
absent, and Eq.~(\ref{diffmig}) is exactly the classical 
continuous description of ion diffusion and migration.

So far, the mean-field equations are quite similar to the ones governing the
evolution of multi-component alloys. However, two important new elements
have now to be taken into account: (i) how to calculate the electric
potential in the mean-field model, and (ii) the electrochemical reactions at
the electrodes.

\subsection{The Poisson problem}

The electric potential has to be obtained, as before, by resolving Poisson's
equation. However, some additional considerations are necessary, because
it turns out that the treatment of the boundary conditions at the
electrode surfaces becomes non-trivial in the mean-field context. 
To understand this, it is useful to start with some comments on the 
consequences of the mean-field approximation. The parameters of the 
model that control the phase diagram are the various interaction
energies and the temperature. Obviously, we want to create and maintain two
distinct phases, namely the metallic electrodes and the electrolyte. Hence,
the interaction energies need to be chosen such that phase separation
occurs. As usual for this kind of lattice models, there exists a critical
temperature for phase separation. Close below the critical point, the
equilibrium concentrations of the two phases are close to each other, that
is, there are many metal particles in the electrolyte phase, and vice versa.
The concentration of these minority species decreases with temperature. In
the original lattice gas model, it is then possible for low enough
temperatures to identify the geometry of the bulk phases with the connected
clusters of metal and electrolyte species.

In the mean-field approximation, the concentration variables (occupation
probabilities) are continuous, and each species has a non-vanishing
concentration at each site. This means that the electrode always contains
small quantities of solvent and ions, and the electrolyte contains a small
quantity of metal. Of course, these concentrations can be made arbitrarily
small by lowering the temperature. However, we then face another difficulty.
The interface between the two bulk phases, which was the sharp boundary
between connected clusters in the discrete model, is now diffuse with a
characteristic thickness that depends on the temperature. For low
temperatures, this thickness becomes smaller than the lattice spacing, which
leads to strong lattice effects on static and dynamic properties of the
interface \cite{Cahn60,Kessler90,Plapp97b}: the surface tension and mobility
of the interface depend on its position and orientation with respect to the
lattice, and for very low temperatures the interface can be entirely pinned
in certain directions. As a consequence, we have to make a compromise in
choosing the temperature: it must be low enough to obtain reasonably low
concentrations of the minority species, but high enough to avoid lattice
pinning and to obtain a reasonably diffuse interface (that extends over a
few lattice sites).

This has consequences for the resolution of Poisson's equation. The boundary
conditions of constant electric potential in the electrodes are easy to
impose in a model with sharp boundaries. For diffuse interfaces, we have to
decide where the electrodes end. Another way to state the problem is to
remark that the existence of an electric field in the electrolyte creates
surface charges in the conductor that are localized at the surface. In a
system with diffuse interfaces, this surface charge is ``smeared out'' over
the thickness of the interface, and we need a method to determine this
charge distribution.

We solve these problems by the introduction of electrons that are free to
move within the metallic electrodes and solve the Poisson equation for all
the charges, including electrons. More precisely, we denote by $p_{\mathbf{k}
}^{e}$ the deviation from the neutral state expressed in electrons per site.
Hence, $p_{\mathbf{k}}^{e}>0$ corresponds to an excess of electrons, $p_{%
\mathbf{k}}^{e}<0$ to an electron deficit. Their chemical potential is
defined by 
\begin{equation}
\widetilde{\mu }_{\mathbf{k}}^{e}=E_{F}+q^{e}V_{\mathbf{k}}+
\frac{p_{\mathbf{k}}^{e}}{\mathcal{D}(E_{F})},  \label{eq12a}
\end{equation}
$\mathcal{D}(E_{F})$ being the density of electronic states at the Fermi
level $E_{F}$ of the metal, and $q^{e}=-e$. This corresponds to the
screening approximation of Thomas-Fermi \cite{Chaz}. The electronic current
is then written as an electronic mobility times the discrete gradient of the
chemical potential, 
\begin{equation}
\widetilde{J}_{\mathbf{k,k+a}}^{e}=
 -\widetilde{M}_{\mathbf{k,k+a}}^{e}\mathfrak{D}_{\mathbf{a}}
  \widetilde{\mu }_{\mathbf{k}}^{e},  \label{eq12b}
\end{equation}
and the time evolution of the excess electron concentration is 
\begin{equation}
\frac{\partial p_{\mathbf{k}}^{e}}{\partial t}=
  -\sum_{\mathbf{a}}\widetilde{J}_{\mathbf{k,k+a}}^{e} .
\end{equation}
Note that the choice of this dynamics is somewhat arbitrary; it could be
replaced by any other process that leads to an equilibrium of the electron
distribution with the ionic charge distribution on a time scale that is much
faster than the characteristic time of evolution of the other species.

One important feature is that the electrons must remain in the electrodes;
otherwise, a ``leakage'' current may occur. We have tested two different
possibilities to achieve this goal. We can either consider that the
electrons are in a potential well in the metallic regions such that they
remain confined in the metal, or we can suppose that their mobility vanishes
outside the metallic regions. Here, we have chosen the second possibility
and written the mobility in the form, 
\begin{equation}
\widetilde{M}_{\mathbf{k,k+a}}^{e}=\frac{w^{e}}{kT} f(p_{\mathbf{k}
}^{0})f(p_{\mathbf{k+a}}^{0}),  \label{eq13b}
\end{equation}
where $w^e$ is a constant frequency prefactor, and $f$ is an interpolation
function that is equal to $1$ for large metal concentrations and falls to
zero for low metal concentrations. With this choice, the electronic jump
probability is important only if nearest neighbor sites $\mathbf{k}$ and $%
\mathbf{k+a}$ have a large enough probability to be occupied by metallic
atoms. We have used for $f$ 
\begin{equation}
f(p)=\frac{\tanh [(p-p_{c})/\xi ]+\tanh [p_{c}/\xi ]}{\tanh [(1-p_{c})/\xi
]+\tanh [p_{c}/\xi ]},  \label{eq13c}
\end{equation}
a monotonous function that varies from $0$ when $p=0$ to $1$ for $p=1$, with
a rapid increase through an interval in $p$ of order $\xi$ centered around
some concentration $p_{c}$ that is reminiscent of a percolation threshold.
This interpolation is motivated by the fact that the metallic region must
be dense enough to be connected and thus allow the electrons to propagate.

To determine the electrostatic potential, we solve the mean-field version of
Poisson's equation, including the new contribution from the electrons, 
\begin{equation}
\sum_{\mathbf{a}}V_{\mathbf{k+a}}-4V_{\mathbf{k}}=-\frac{a^{2-d}}{\epsilon }%
\sum_{\alpha =+,-,e}q^{\alpha }p_{\mathbf{k}}^{\alpha }.  \label{eqpoisson}
\end{equation}
This equation is now solved in the whole system, including the electrodes.
Note that we still use a constant permittivity $\epsilon$; however, the
phase-dependent mobility of the electrons makes the resistivity in the
electrolyte much higher than in the metal.

This method provides a fast way to calculate the surface charges on the
electrodes. As will be shown below, it works perfectly well at equilibrium.
However, in out-of-equilibrium simulations, a problem appears on the side of
the anode where the metal is dissolved. Since the mobility rapidly decreases
with the metal concentration, electrons present on the metallic site before
dissolution may be trapped in the electrolyte, leading to spurious
electronic charges in the bulk. We have solved this problem in a
phenomenological way by adding a term to the evolution equation for the
electrons that relaxes the electronic charge to zero in the electrolyte, 
\begin{equation}
\frac{\partial p_{\mathbf{k}}^{e}}{\partial t}=-\sum_{\mathbf{a}}\widetilde{J%
}_{\mathbf{k,k+a}}^{e}-w^{e}[1-f_{r}(p_{\mathbf{k}}^{0})]p_{\mathbf{k}}^{e},
\label{eqdpedt}
\end{equation}
where $f_{r}(p_{{}}^{0})$ is the same interpolation function as $f$, but
with different parameters $\xi _{r}$ and $p_{c,r}$. With a convenient choice
of these parameters, ``electron relaxation'' occurs only in the liquid.

\subsection{Electron transfer}

On the electrode surfaces, electron transfer takes places. More
precisely, metallic cations $M^+$ located in the electrolyte may 
receive an electron from a neighboring metallic site and be reduced; 
in turn, metal atoms in contact with the electrolyte may reject an 
electron to a neighboring metallic site and become an ion, 
\begin{equation}
M^{+}(\mathbf{k)}+e^{-}(\mathbf{k+a)}\rightleftharpoons M^{0}(\mathbf{k)}.
\label{eq5}
\end{equation}
The direction of the transfer depends on the relative magnitude of the
electrochemical potentials of the involved species. Reduction of cations on
a site $\mathbf{k}$ of the cathode appears when 
\begin{equation}
\widetilde{\mu }_{\mathbf{k}}^{+}+\widetilde{\mu }_{\mathbf{k+a}}^{e}>%
\widetilde{\mu }_{\mathbf{k}}^{0},  \label{eq16}
\end{equation}
otherwise, the metal is oxidized. Consequently, we define $\sigma _{\mathbf{%
k,k+a}}$ as the current of electronic charges from $\mathbf{k+a}$ to $%
\mathbf{k}$ (current of positive charges from $\mathbf{k}$ to $\mathbf{k+a}$
) reducing the cations on site $\mathbf{k}$ (resp. electronic current issued
from the oxidation of the metal) via a corresponding elimination (resp.
creation) of electrons on site $\mathbf{k+a}$. Following \cite{Schm}, we can
write the reaction rate, 
\begin{equation}
\sigma _{\mathbf{k,k+a}}=w_{\mathbf{k,k+a}}^{*}\left( \exp \frac{\widetilde{
\mu }_{\mathbf{k}}^{+}+\widetilde{\mu }_{\mathbf{k+a}}^{e}}{kT}-\exp \frac{%
\widetilde{\mu }_{\mathbf{k}}^{0}}{kT}\right) .  \label{eq14}
\end{equation}
This corresponds to an activated electronic charge transfer between the
metal surface and the nearest neighboring cation. The total reduction rate
on site $\mathbf{k}$ is the sum of all the reaction paths $\sum_{\mathbf{a}%
}\sigma _{\mathbf{k,k+a}}$. The coefficient $w_{\mathbf{k,k+a}}^{*}$ can be
determined by comparison with the mesoscopic theory of Butler-Volmer (see
the Appendix).

The currents $\sigma _{\mathbf{k,k+a}}$ have non negligible values only on
the interfaces. However, as discussed before, there are small concentrations
of ions in the metal and metal atoms in the electrolyte; this would lead to
undesirable contributions of the reaction term inside the bulk phases.
Therefore, we use the same interpolation function $f(p)$ as for the electron
mobility and write 
\begin{equation}
w_{\mathbf{k,k+a}}^{*}=w^{*}\left( 1-f(p_{\mathbf{k}}^{0})\right) f(p_{%
\mathbf{k+a}}^{0})\;,  \label{eq15}
\end{equation}
where $w^{*}$ is again a constant frequency factor. In this way, the
transfer is localized around the metal-electrolyte interface, where occupied
metallic sites and electrolyte sites are neighbors. To completely suppress
the electrochemical processes in the bulk phases, we set the transfer
current to zero when the product of the two interpolation prefactors is
smaller than $10^{-4}$.

\subsection{Summary}

After all the different pieces are combined, the complete set of equations is 
\begin{equation}
\frac{\partial p_{\mathbf{k}}^{+}}{\partial t}=-\sum_{\mathbf{a}}\widetilde{J%
}_{\mathbf{k,k+a}}^{+}-\sum_{\mathbf{a}}\sigma _{\mathbf{k,k+a}},
\label{eq4}
\end{equation}
\begin{equation}
\frac{\partial p_{\mathbf{k}}^{0}}{\partial t}=-\sum_{\mathbf{a}}\widetilde{J%
}_{\mathbf{k,k+a}}^{0}+\sum_{\mathbf{a}}\sigma _{\mathbf{k,k+a}},
\label{eq6}
\end{equation}
\begin{equation}
\frac{\partial p_{\mathbf{k}}^{-}}{\partial t}=-\sum_{\mathbf{a}}\widetilde{J%
}_{\mathbf{k,k+a}}^{-},  \label{eq7}
\end{equation}
\begin{equation}
\frac{\partial p_{\mathbf{k}}^{s}}{\partial t}=-\sum_{\mathbf{a}}\widetilde{J%
}_{\mathbf{k,k+a}}^{s},  \label{eq7a}
\end{equation}
\begin{equation}
\frac{\partial p_{\mathbf{k}}^{e}}{\partial t}=-\sum_{\mathbf{a}}\widetilde{J%
}_{\mathbf{k,k+a}}^{e}-w^{e}[1-f_{r}(p_{\mathbf{k}}^{0})]p_{\mathbf{k}%
}^{e}-\sum_{\mathbf{a}}\sigma _{\mathbf{k+a,k}},  \label{eq8}
\end{equation}
\begin{equation}
\sum_{\mathbf{a}}V_{\mathbf{k+a}}-4V_{\mathbf{k}}=-\frac{a^{2-d}}{\epsilon }%
\sum_{\alpha =+,-,e}q^{\alpha }p_{\mathbf{k}}^{\alpha }.  \label{eq9}
\end{equation}
The local concentration of the species $M^{0}$ and $M^{+}$ is modified by
transport (diffusion and migration in the electric field) and by the
chemical reaction; for the other species $A^{-}$ and $S$, only transport is
present. The reaction term is active only in the solid-electrolyte interface
and has of course an opposite sign in Eqs.~(\ref{eq4}) and (\ref{eq6}), and
no contributions in Eqs.~(\ref{eq7}) and (\ref{eq7a}).

Equations (\ref{eq4}) through (\ref{eq8}) are integrated in time by a simple
Euler scheme (with constant or variable timestep). Two methods are used to
solve Poisson's equation. Equation (\ref{eq9}) can either be solved for each
timestep using a conjugate gradient method, or it can be converted into a
diffusion equation with a source term and solved by simple relaxation; the
second method generally is computationally faster and converges well for
interfaces that move slowly enough.

\section{Numerical tests}

\subsection{Chemical equilibrium}

The complete EMFKE model involves a considerable number of parameters. In
particular, we need to make a choice for all the short-range interaction
energies. They determine the phase diagram of the four-component system
that consists of metal, solvent, anions, and cations. 
To obtain meaningful results, the model has to be used
not too far from a two-phase equilibrium between a metal-rich electrode and
a liquid phase rich in solvent and ions. For an arbitrary choice of
interaction energies, obtaining the equilibrium concentrations for all of
the species in the two phases is not a trivial task. The conditions for a
two-phase equilibrium are (i) equal electrochemical potentials for each
independent species (four in our case), and (ii) equal grand potential. This
yields five conditions for eight unknowns (four equilibrium concentrations
in each phase); consequently, three degrees of freedom remain that may be
fixed, for example the concentrations of ions and metal in the liquid. This
constitutes a set of five coupled nonlinear equations for the remaining
unknowns.

To simplify the task of finding an equilibrium configuration, we start from
a simpler system, namely a mixture of metal, solvent, and vacancies, that is
equivalent to a binary alloy with vacancies (ABv). For a symmetric
interaction matrix, $\epsilon^{00}=\epsilon^{ss}$, the phase diagram can be
determined analytically \cite{PlappPRL}. A two-phase equilibrium exists with 
$p^s_{liq}=p^0_{sol}$ and $p^0_{liq}=p^s_{sol}$. The equilibrium
concentrations can be expressed in terms of the new variables $%
P=p^s_{liq}+p^0_{liq}=p^s_{sol}+p^0_{sol}$ and $%
Q=p^s_{liq}-p^0_{liq}=p^0_{sol}-p^s_{sol}$ and the reduced interaction
energy $\bar\epsilon=(\epsilon^{00}+\epsilon^{ss}-2\epsilon^{0s})/2$ as 
\begin{equation}
P=Q / \tanh \left(\frac{z\bar\epsilon Q}{2kT}\right),
\end{equation}
where $z$ is the coordination number ($z=4$ for a square lattice in two
dimensions). For the rest of the section, we choose $\epsilon^{00}/kT=%
\epsilon^{ss}/kT=1$, $\epsilon^{0s}=0$, and $Q=0.9$, which gives 
$p^s_{liq}=p^0_{sol}=0.92528$, $p^0_{liq}=p^s_{sol}=0.02528$, and 
$p^v_{liq}=p^v_{sol}=0.04944$.

Next, this equilibrium is perturbed by the addition of ions. If the
interaction energies are taken identical for both species of ions, no
separation of charge occurs at equilibrium and in the absence of an applied
voltage, and we can for the moment omit the Poisson equation. We take $%
\epsilon^{+s}/kT=\epsilon^{-s}/kT=1$ and $\epsilon^{++}=\epsilon^{--}=%
\epsilon^{+0}=\epsilon^{-0}=0$. This means that the ions are attracted by
the solvent, and that the energy density of the liquid does not change upon
addition of ions. As long as the equilibria of the main components are not
appreciably modified, we expect an equilibrium distribution of the ions that
satisfies 
\begin{equation}
{\frac{p^{\pm}_{sol}}{p^{\pm}_{liq}}} = 
   \exp\left(-{\frac{z\epsilon^{\pm s}Q}{kT}}\right).
\label{partition}
\end{equation}

We tested this prediction by performing simulations of one-dimensional half
cells of length 40, in which 20 lattice sites of solid are in contact with
20 lattice sites of liquid. In each phase, all concentration values are
initially set to the expected equilibrium values. On the solid side, we
apply no-flux boundary conditions, whereas at the liquid side the
concentrations are kept constant. This corresponds to a finite-size isolated
electrode plunged in an infinite solution. The electrochemical reaction is
suppressed by setting $w^\star=0$; all other frequency factors $w^\alpha$
are set to $1$. The equations are integrated with a fixed timestep of $%
\Delta t=1$.

\begin{figure}[tbp]
\begin{center}
\includegraphics[width=7cm]{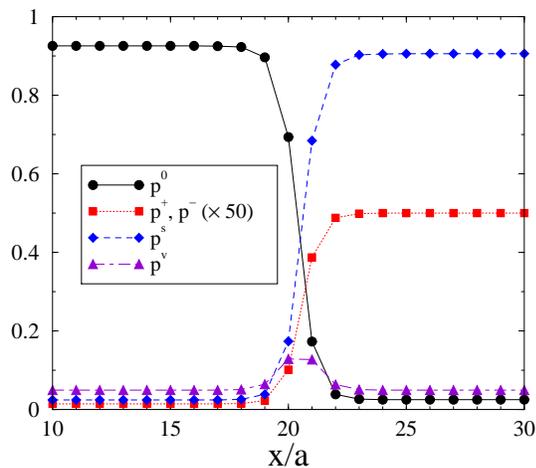}
\end{center}
\caption{Concentration profiles obtained from the relaxation of a step
profile (20 sites of solid in contact with 20 sites of electrolyte) with ion
concentration in the liquid of $p^\pm_{liq}=0.01$. The other parameters are
given in the text.}
\label{figprof}
\end{figure}

\begin{figure}[tbp]
\begin{center}
\includegraphics[width=7cm]{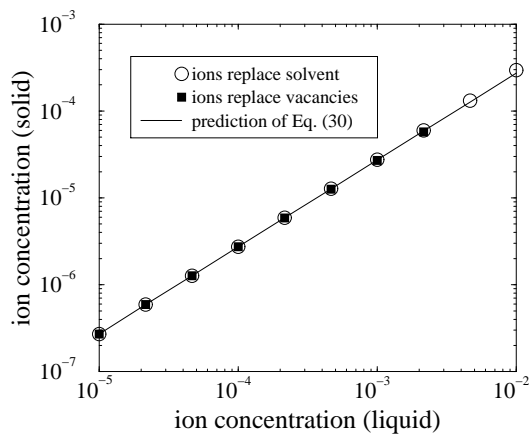}
\end{center}
\caption{Ion concentration in the solid $p^\pm_{sol}$ in equilibrium with a
liquid of fixed ion content $p^\pm_{liq}$.}
\label{figconcs}
\end{figure}

Two ways of adding ions were tested. In the first case, $p^v_{liq}$ is kept
constant and a small percentage of solvent is replaced by ions. Within a
few $10^6$ timesteps, the chemical potential differences across the cell
fall below $10^{-9}$ and the evolution virtually stops. The concentration
profiles obtained for $p^+_{liq}=p^-_{liq}=0.01$ are shown in Fig.~\ref
{figprof}. In Fig.~\ref{figconcs}, we plot the ion concentration in the
solid versus the ion concentration in the liquid. For low concentrations,
Eq.~(\ref{partition}) is well satisfied. For concentrations larger than $%
\approx 10^{-3}$, deviations appear because the concentrations of metal,
solvent and vacancies are shifted. However, even for $p^\pm_{liq}=0.01$ this
deviation is less than 5\%.

In the second case, ions were added in replacement of vacancies ($p^s_{liq}$
and $p^0_{liq}$ were kept constant). Here, the solvent-metal equilibrium is
shifted more rapidly due to the strong dependence of the chemical potentials
on the vacancy concentration. For ion concentrations larger than $0.003$,
the interface never relaxes, but starts to grow by incorporating metal that
is transported to the interface from the solution by chemical diffusion. Of
course, the fact that the system does not reach equilibrium is due to the
fact that we have fixed one degree of freedom too much, namely all four
concentrations in the liquid. Consequently, even the solutions found by the
first method are not, strictly speaking, equilibrium solutions; however,
they are sufficiently close to equilibrium for all practical purposes and
will hence be used as initial conditions in the following.

\subsection{Poisson equation and screening: perfectly polarizable electrode}

To test our method for creating interface charges, we started from an
equilibrated half cell as calculated in the preceding subsection, added a
potential difference between electrode and solution, and opened the metal
side of the cell for a current of electrons (but not of other species). 
Since the electron transfer frequency was kept to zero ($w^\star=0$),
and hence no electrochemical processes can take place, this situation
corresponds to a grounded perfectly polarizable electrode.

\begin{figure}[tbp]
\begin{center}
\includegraphics[width=7cm]{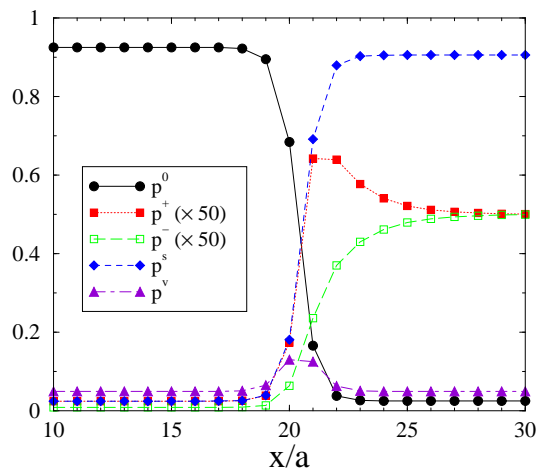}
\end{center}
\caption{Concentration profiles with the same parameters as in Fig.~\ref
{figprof}, but with a potential difference $\Delta V = -0.5 kT/e$ between
metal and solution.}
\label{figlayer}
\end{figure}

For simplicity, we have restricted our attention here to the case of 
monovalent ions ($q^+=e$, $q^-= -e$). In our tests, we want the
thickness of the charged double layer to be a few lattice constants. This
can be achieved by choosing the dielectric constant $\epsilon=0.05e^2/(akT)$.
Furthermore, we choose for the density of states at the Fermi level 
${\mathcal{D}(E_{F})}=1000/kT$. Since $kT{\mathcal{D}(E_{F})}\gg1$, the
electrochemical potential of the electrons depends only weakly on the
surface charge. The parameters of the interpolation functions are fixed to $%
p_c=0.5$, $\xi=0.1$ for the electron mobility (this leads to an electron
mobility that is 8 orders of magnitude smaller in the liquid than in the
electrodes) and $p_{c,r}=0.03$, $\xi_r=0.001$ for the electron relaxation
(see Eq.~(\ref{eqdpedt}); this corresponds to an interpolation
function with a sharp increase at a concentration slightly
larger than the equilibrium metal concentration in the liquid). The
resulting interface profile for $p^\pm_{liq}=0.01$ and a potential
difference of $\Delta V = V_{metal}-V_{liquid}=-0.5 kT/e$ is plotted in Fig.~%
\ref{figlayer}. As expected, the cations are attracted to the electrode,
whereas the anions are repelled.

\begin{figure}[tbp]
\begin{center}
\includegraphics[width=7cm]{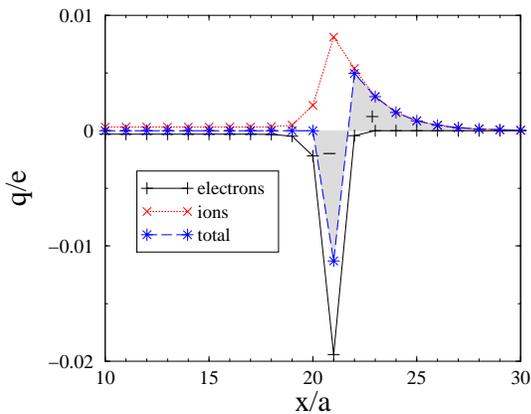}
\end{center}
\caption{Charge distribution ($q = q^\alpha p^\alpha$ for species $\alpha$)
across the interface of Fig.~\ref{figlayer}.}
\label{figcharges}
\end{figure}

In Fig.~\ref{figcharges}, we show the charge distribution that corresponds
to the profile of Fig.~\ref{figlayer}. Several features are noteworthy. The
small homogeneous ion concentrations inside the electrodes are now different
for positive and negative ions. This is a direct consequence of the global
equilibrium: since the electrochemical potentials are constant throughout
the whole system and the ion concentrations in the liquid are fixed, the ion
concentrations in the solid are multiplied by factors $\exp(-q^\alpha\Delta
V /kT)$ with respect to the reference state $\Delta V=0$. The resulting
``background charge'' in the bulk electrode is exactly compensated by the
electrons that have flown in from the metal side of the system. Since the
resulting total charge inside the bulk electrode is zero, the potential is
constant, which was the original purpose of introducing the electrons.
Inside the diffuse interface between electrode and solution, we find a
surface charge that is located almost exclusively on one site.

\begin{figure}[tbp]
\begin{center}
\includegraphics[width=7cm]{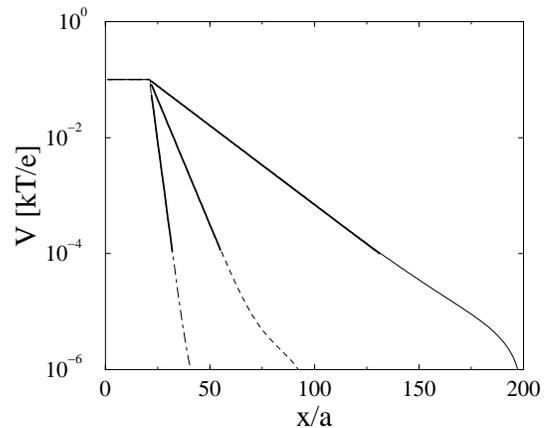}
\end{center}
\caption{Potential $V$ as a function of $x$ for $\Delta V = 0.1 kT/e$ and $%
p_{liq}^\pm$=0.0001 (solid line), $p_{liq}^\pm$=0.001 (dashed line), and $%
p_{liq}^\pm$=0.01 (dash-dotted line). The thick line segments are
exponential fits ($V(x)=\Delta V \exp[-\kappa(x-x_{int})]$) that yield $%
\kappa=0.0626$, $0.199$, and $0.622$ for the inverse of the Debye length.
The predictions of Eq.~(\ref{debye}) are $0.0632$, $0.200$, and $0.632$,
respectively. The sizes of the simulated cells were $200$, $100$, and $50$
lattice spacings.}
\label{figexp}
\end{figure}

In the liquid, we find the diffuse charged layer that is predicted 
by the classic sharp-interface model of Gouy and Chapman. For an
electrolyte that has singly charged ions, $q^+ = - q^- = e$, 
the potential and the deviations of the concentrations from the 
bulk liquid values all decay exponentially with the distance from
the interface. For example, $V(x)=\Delta V \exp[-\kappa(x-x_{int})]$ 
with a decay constant 
\begin{equation}
\kappa = \sqrt{\frac{2e^2p_{liq}^\pm}{\epsilon kT}},  \label{debye}
\end{equation}
the inverse of the Debye screening length, and $x_{int}$ the position
of the (sharp) interface. In Fig.~\ref{figexp}, we show the potential 
profile across a cell subjected to a potential difference of 
$\Delta V = 0.1 kT/e$ for different ion concentrations. We
obtain values for $\kappa$ by an exponential fit of $V(x)$, using 
$x_{int}$ (the position of the ``sharp interface'' extrapolated 
from the far field) as an adjustable parameter. The results are in
good agreement with the predictions of Eq.~(\ref{debye}). The potential
follows an exponential law over several orders of magnitude. However, 
when the potential approaches its reference value in the liquid,
deviations from the exponential behavior occur. This is due to the
fact that the buildup of the ion boundary layer at the interface also
modifies the chemical potentials for solvent and metal by small amounts.
Since the electrochemical potential of the ions depends on the solvent and
vacancy concentrations, a complete equilibrium solution must contain these
effects. Close to the interface, they are negligible in
comparison to the contribution of the electric potential;
only far from the interface, when $V-V_{liquid}$ becomes of the
same order as the small shifts in the chemical potentials, the
deviations from an exponential become appreciable. Since this 
occurs for fractions of the potential drop that are less than 
$10^{-3}$ of $\Delta V$, these effects can safely be neglected 
in the further analysis.

The resulting total charge distribution has the expected double-layer
structure, with a much sharper decay in the electrode than in the liquid. We
have checked that the integral of the total charge through the interface is
zero to numerical precision.

\subsection{Electrochemical equilibrium: Nernst law}

The electrochemical equilibrium between electrodes and electrolyte depends
on the choice of the Fermi energy $E_F$. Indeed, consider a chemical
equilibrium in the absence of an electric potential such as the one shown in
Fig.~\ref{figprof}. In this state, the electrochemical potential of the
ions is constant across the cell. If we fix $E_F$ to be exactly equal to the
difference of the chemical potentials of metal and cations, 
\begin{equation}
E_F^0 = \mu^0_{eq} - \mu^+_{eq},  \label{ef0def}
\end{equation}
the reaction currents $\sigma$ are strictly zero everywhere as long as the
potential remains constant. Therefore, the original profile remains an
equilibrium state for arbitrary transfer frequency $w^\star$. In contrast,
if $E_F$ is set to a value different from $E_F^0$, the electron transfer 
starts and drives the system to a new equilibrium that exhibits a 
potential difference.

The final potential difference will be close to $E_F - E_F^0$. This can be
deduced as follows. For fixed concentrations and electric potential in the
liquid, the electrochemical potentials of the ions and the metal are fixed.
Therefore, the only way to achieve equilibrium is to adjust the electric
potential in the electrode such that the two terms in Eq.~(\ref{eq14})
balance. Now consider the electrochemical potential of the electrons. The
excess electronic charge $p^e$ is always small, $|p^e|\ll e\Delta V/kT$,
even at the surface, as can be seen in Fig.~\ref{figcharges}. Since, in
addition, we have chosen $\mathcal{D}(E_F) kT \gg 1$, $\widetilde \mu^e$ is
only weakly dependent on the local electron density. Neglecting the last
term in Eq.~\ref{eq12a}, the electron transfer rate becomes zero when $%
E_F-e\Delta V = E_F^0$.

We have checked this prediction by performing simulations in isolated cells:
the concentrations of all species and the potential were kept fixed at the
liquid side, and the system was closed at the solid side for particles and
electrons, thus enforcing zero electrical current. For $p^\pm_{liq}=0.01$,
Eq.~(\ref{ef0def}) yields $E_F^0/kT = 4.447514$. For $E_F = E_F^0$, the
system remained at constant potential, whereas for $E_F \neq E_F^0$, the
predicted potential difference developed up to an error of less than $10^{-5}
$. Since all of the charges can be removed from the system by applying an 
\emph{external} potential difference that is just the negative of $%
(E_F-E_F^0)/e$, the choice of $E_F$ determines the potential of zero charge
(PZC) of the electrochemical interface.

\begin{figure}[tbp]
\begin{center}
\includegraphics[width=6.5cm]{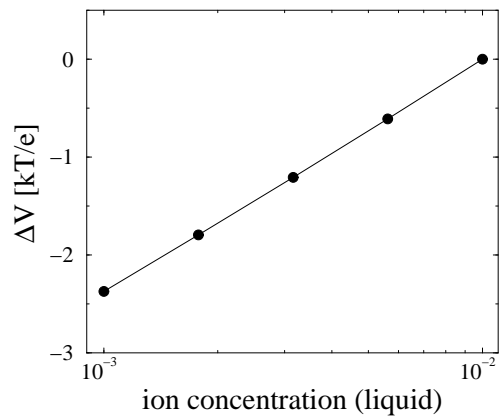}
\end{center}
\caption{Equilibrium potential difference through an isolated cell versus
ion concentration in the liquid for fixed $E_F=4.447514\, kT$. The solid
line corresponds to Eq.~(\protect\ref{nernst}).}
\label{fignernst}
\end{figure}

Of course, the value of $E_F^0$ as defined above depends on the ion
concentration in the liquid. Therefore, for fixed $E_F$, there is a
well-defined ion concentration $p^\pm_{ref}$ for which the potential
difference is zero. When the concentration in the liquid is varied, the
potential difference follows Nernst's law, 
\begin{equation}
\Delta V = {\frac{kT}{e}} \ln {\frac{p^\pm}{p^\pm_{ref}}} .  \label{nernst}
\end{equation}
In Fig.~\ref{fignernst}, we plot $\Delta V$ vs $p^\pm_{liq}$ for $E_F^0/kT =
4.447514$ (corresponding to $p^\pm_{ref}=0.01$). The results of our model
are in perfect agreement with Eq.~(\ref{nernst}).

\subsection{Growth}

To investigate growth and dissolution, we use one-dimensional cells with a
metallic electrode at both ends. The system is closed on both sides for all
species (ions, metal, solvent) but open for electrons. To drive the
interfaces, we apply a fixed potential difference $\Delta V$ across the cell.
For $e\Delta V/kT$ ranging from 1 to 10 and a cell of size 100 with two
electrodes of thickness 10, the system reaches a steady state within $%
5\times 10^6$ timesteps ($\Delta t = 1$). Within $2\times 10^7$ timesteps,
the interface advances between 3 and 15 sites, depending on the voltage. We
measure the ionic current in the center of the cell by time-averaging over
the second half of the runs.

Given the rapid variations of ion concentrations and electronic charges
through the interface (see for example Fig.~\ref{figcharges}), one could
expect strong lattice pinning effects \cite{Cahn60,Kessler90,Plapp97b}. We
have checked that quantities such as the electronic surface charge, the
total transfer rate (that is, the transfer currents $\sigma _{\mathbf{k,k+a}}
$ summed up through the interface) and the ion current indeed do vary as the
interface advances through the lattice. However, for the parameters chosen
here, the amplitudes of these lattice oscillations never exceeded a few
percent.

The choice of the various time constants does not influence the final
results for equilibrium states. In contrast, for growth simulations, they
have to be fixed in order to achieve the desired physical conditions. In
particular, this is true for the various jump rates and the electron
transfer frequency. The electric current in the electrolyte is carried by
the mobile ions. As mentioned above, the buildup of the ionic boundary
layers shifts the chemical potential of metal, solvent, and vacancies as
well. Since, in our model, the concentration of metal in the electrolyte is
comparable to the ion concentrations, this leads to neutral diffusion
currents that are unphysical. To lower their magnitude, the jump frequency
for the metal was chosen much smaller than for the ions: $w^+=w^-=1$, $%
w^0=10^{-3}$.

\begin{figure}[tbp]
\begin{center}
\includegraphics[width=7cm]{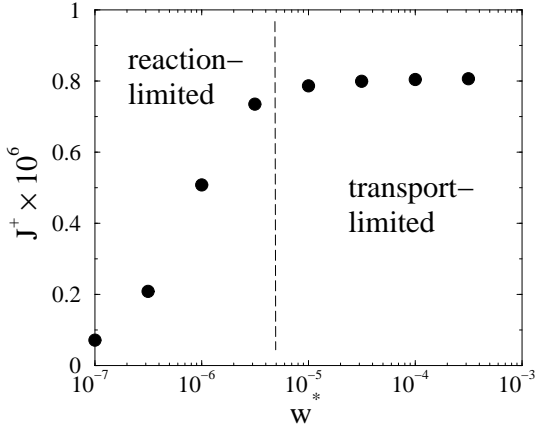}
\end{center}
\caption{Ion current in the center of the cell as a function of the electron
transfer frequency $w^\star$ for fixed voltage $\Delta V = 10\,kT/e$.}
\label{figjvsomeg}
\end{figure}

In Fig.~\ref{figjvsomeg}, we plot the ionic current in the center of the
cell as a function of the transfer frequency $w^\star$ at fixed voltage of
10 $kT/e$. For low transfer frequencies, the growth is limited by the
reaction kinetics at the interface, and the current strongly depends on the
value of $w^\star$. For increasing $w^\star$, the growth becomes limited by
transport in the bulk, and the current is almost independent of $w^\star$.
We are mostly interested in the latter regime. Therefore, we fix in the
following $w^\star=10^{-4}$.

\begin{figure}[tbp]
\begin{center}
\includegraphics[width=7cm]{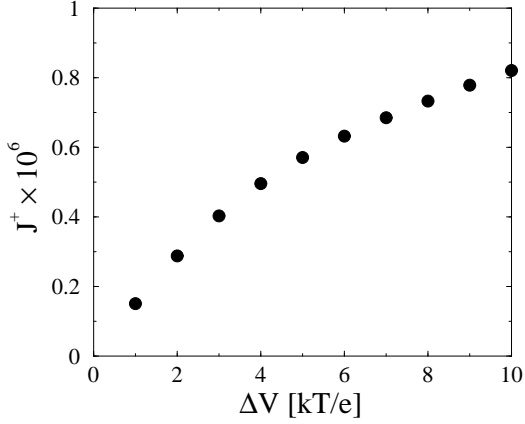}
\end{center}
\caption{Ion current in the center of the cell as a function of the driving
voltage for fixed $w^\star = 10^{-4}$.}
\label{figjvsv}
\end{figure}

\begin{figure}[tbp]
\begin{center}
\includegraphics[width=7cm]{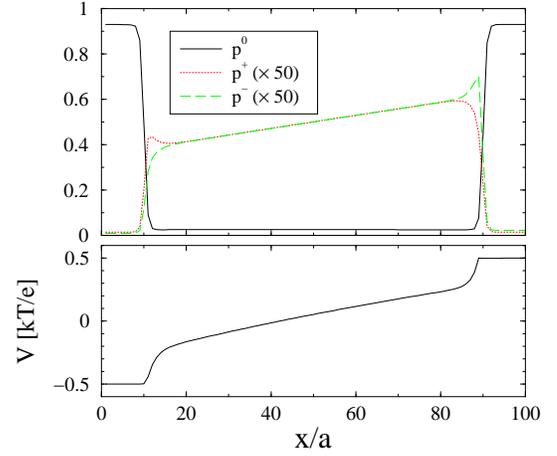}
\end{center}
\caption{Concentration profiles (top) and potential profile (bottom) across
a 100-site cell with initial ion concentration of 0.01 subjected to a
potential difference of $\Delta V = kT/e$. The snapshot was taken at $%
t=2\times 10^7$.}
\label{figv1}
\end{figure}

\begin{figure}[tbp]
\begin{center}
\includegraphics[width=7cm]{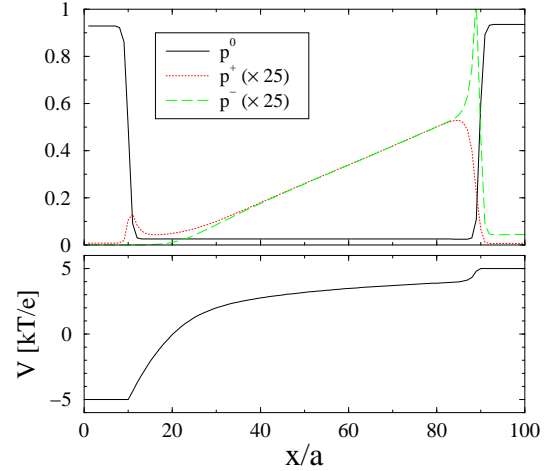}
\end{center}
\caption{Same as Fig.~\ref{figv1}, but for a potential difference of $\Delta
V = 10\,kT/e$.}
\label{figv10}
\end{figure}

The current-voltage curve for our model cell is shown in Fig.~\ref{figjvsv}.
It is strongly nonlinear. Indeed, two very different regimes are covered by
these simulations. This can be appreciated when looking at the ion and
potential profiles in Figs.~\ref{figv1} and \ref{figv10}. For $\Delta V =
kT/e$ (Fig.~\ref{figv1}), the ion concentrations remain of the same order of
magnitude as the initial concentration ($p^\pm=0.01$), and except for the
two double layers close to the interfaces, the liquid is neutral. An
important part of the potential drop occurs in these double layers; in the
bulk electrolyte, the potential profile is smooth and almost linear. In
contrast, for $\Delta V = 10 kT/e$ (Fig.~\ref{figv10}), the neighborhood of
the cathode has been completely depleted of the anions ($p^-\sim 10^{-5}$ at
the cathode), and a charged zone extends well beyond the thickness of the
equilibrium double layers. Most of the potential drop occurs close to the
cathode. Since the conductivity in the space-charge zone is lowered due to the 
low ion concentration, a considerable increase in the overall voltage 
$\Delta V$ leads only to a moderate increase of the current, as seen 
in Fig.~\ref{figjvsv}. All of these findings are in good agreement with 
the macroscopic one-dimensional calculations of Chazalviel \cite{Chaza90}. 
According to this work, the extended space charge is crucial for the 
emergence of ramified growth: the strong electric field close to the 
surface leads to an instability of the flat front, and one-dimensional 
calculations become invalid.

\subsection{Two-dimensional simulations: dendritic growth}

We present now an example for a preliminary simulation of 
a two-dimensional sample. The purpose is to show that our
model can indeed lead to the emergence of dendritic 
structures; however, the conditions that we can simulate
are far from typical experimental situations. The main reason 
is that in experiments, the branches of ramified aggregates 
have typically a thickness in the micron range, whereas the 
lattice of our model represents a crystal lattice with spacing 
of the order \AA ngstr\"oms. It is clear that huge simulation cells 
would be necessary to observe instabilities and ramified growth
on realistic scales. To obtain a computationally tractable
problem, we have to work with unrealistically high driving forces,
since this is known to reduce the characteristic scales of
branched growth structures. Therefore, we use a much higher
driving potential than in the previous simulations,
namely 100 $kT/e$. This corresponds to about 2.5 V at room
temperature, which is a fairly typical value; however, this
potential difference is applied through a cell that is, as
before, 100 lattice sites long, which corresponds to a
length of a few nanometers. Therefore, the electric fields
are much higher in our simulation than in reality. 

Under these conditions, the behavior of the moving interface
is quite sensitive to the model parameters, and in particular
to the frequency factors for the electron transfer, $w^\star$,
and the metal jump frequency $w^0$ (as before, we take $w^+=w^-=1$
as a reference value). To assure that the growth is still
transport-limited for the higher driving force, $w^\star$ has
to be chosen large enough; for reaction-limited growth, no
morphological instability occurs. The metal jump frequency
has to be chosen carefully. On the one hand, if it is too
high, bumps on the surface are smoothed out too rapidly by
surface diffusion and/or an evaporation-condensation mechanism.
On the other hand, if it is too low, the diffusion of metal
on the solid side of the interface becomes so slow that the
interface profile cannot be maintained, and the metal grows
at a concentration far below its equilibrium value \cite{mobthesis}.

\begin{figure}[tbp]
\begin{center}
\includegraphics[width=8cm]{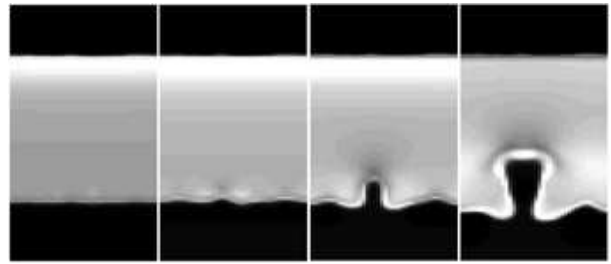}
\end{center}
\caption{Snapshots of the evolution of a $100\times 40$ electrochemical 
cell; the parameters are given in the text. The electrodes are in black; 
the cathode is at the bottom and grows upward (the cell is recentered 
during the simulation). In the electrolyte, the grayscale indicates the
concentration of ions, with white corresponding to the maximal ion
concentration. The white ``contour'' at the electrode surfaces is the
charged double layer; the gray region in front of the dendrite has
been almost completely depleted of ions.}
\label{fig2d}
\end{figure}

In Fig.~\ref{fig2d}, we show an example computed with 
$w^\star=6\times 10^{-3}$, $w^0=10^{-3}$, and $w^e=10^{-3}$.
The cell has a size 40 $\times$ 100 lattice sites, with 
periodic boundary conditions parallel to the interfaces.
The simulation was started from a flat interface, with
random shifts of the metal concentration in the interfaces
to trigger the instability. When one layer of the anode
was dissolved, the whole cell was shifted backward by one
site in order to keep the electrolyte in the center of the
cell. It can be seen that a bump grows on the interface
and develops into a finger-like structure. Other bumps
that initially develop on the interface are screened.
The whole region that surrounds the dendrite is depleted
of ions. An extended charged region forms ahead of the
tip. When the tip gets closer to the anode, the electric
field increases and leads to growth of the metal at
unphysically low concentrations.

\section{Conclusion}

In summary, we have shown here that, starting from a simple microscopic
model, it is possible to build Electrochemical-Mean-Field-Kinetic Equations
(EMFKE) that are able to reproduce qualitatively the behavior of
electrochemical cells. Both the charged double layers present at equilibrium
and the extended space charge that develops during growth are correctly
reproduced. Dendritic structures can be simulated, albeit for unrealistic
parameters. Hence, the EMFKE contain the fundamental ingredients that are 
necessary to simulate dendritic growth by electrodeposition.

Our model shares many common features with a recent phase-field formulation
of electrodeposition \cite{GBW}. The phase-field method, originally
developed in the context of solidification in the 1980s \cite
{Langer86,Collins85,Caginalp86}, is a continuous model of phase transitions
that uses an auxiliary indicator field, the phase field, to distinguish
between the different thermodynamic phases (here, electrodes and
electrolyte). A phenomenological equation of motion for the phase field is
usually derived from a free energy functional. In our EMFKE approach, the
role of the phase field is played by the metal concentration.

Since the phase-field method is phenomenological, it has a certain freedom
of choice for the dynamics of the phase field itself, and can hence avoid
the problems that arise in our model due to the presence of metal in the
electrolyte. However, no direct link to a microscopic model is established,
which is the strength of our approach. Our equations still contain some
phenomenological elements, in particular, the interpolation function for
electron diffusivity and reaction rates. A more realistic modeling of the
processes involving electrons is needed to overcome this limitation. With
this perspective, our approach may constitute a useful link between
microscopic models and phase-field models.

Discussions with J.-N. Chazalviel, V.\ Fleury, and M.\ Rosso are greatly
acknowledged. Laboratoire de Physique de la Mati\`{e}re Condens\'{e}e is
Unit\'{e} Mixte 7643 of CNRS and Ecole Polytechnique.

\appendix

\section{The Butler-Volmer equation}

On the metal-electrolyte interface, the oxido-reduction reaction, 
\begin{equation}
M^{+}(\mathbf{k)}+e^{-}(\mathbf{k+a)} 
\begin{array}{l}
k_{red} \\ 
\rightleftharpoons \\ 
k_{ox}
\end{array}
M^{0}(\mathbf{k)}.  \label{eqA1}
\end{equation}
is characterized by two rates $k_{ox}$ and $k_{red}$. In our simplified
model, the reduction of a cation $M^{+}$ located on a site $\mathbf{k}$ that
is nearest neighbor of a surface site, is carried out by the transfer of an
electron coming from a site $\mathbf{k+a}$ of the electrode: a reduced
metallic atom is then created at site $\mathbf{k}$ on the interface.

\begin{figure}[tbp]
\begin{center}
\includegraphics[width=8cm]{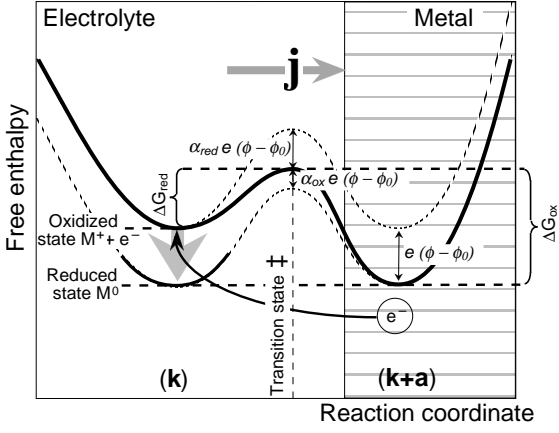}
\end{center}
\caption{Free enthalpy along the reaction path for the electron transfer.
Between the two states $M^{0}$ and $M^{+}+e$, there is a transition 
state $\ddag $ of higher enthalpy. Therefore, the barriers $\Delta G_{red}$
and $\Delta G_{ox}$ determine the reaction rate. It exists an equilibrium
potential $\phi_{0}$ for which $M^{0}$ and $M^{+}+e$ have the same
enthalpy, the two barriers are the same, 
$\Delta G_{red}(\phi _{0})=\Delta G_{ox}(\phi_{0})$, 
and there is a balance between oxidation and reduction. 
When a potential $\phi \neq \phi _{0}$ is applied, the 
relative positions of the states vary, and the barriers 
are modified. In the figure, the solid line is the free 
enthalpy at $\phi\neq\phi_0$, whereas the two dotted lines
are two copies of the equilibrium enthalpy profile, shifted 
to match the two local minima in the non-equilibrium profile.
The definitions of the modified barriers can be read off;
since we have 
$\alpha_{red}e(\phi-\phi_0)+\alpha_{ox}e(\phi-\phi_0)=e(\phi-\phi_0)$,
Eq.~(\protect\ref{eqA4}) follows.}
\label{FigButlerVolmer}
\end{figure}

The Butler-Volmer model (Fig. \ref{FigButlerVolmer}) supposes that there
exists a potential difference $\phi$ between the electrolyte, where the
cations $M^{+}$ can be reduced, and the metal of the electrode, where a
metallic atom $M^{0}$ can be oxidized. In between, there exists an
activation barrier for the redox reaction, and (if we suppose that the
frequency prefactors are the same) the corresponding rates are
\begin{equation}
k_{red}=A\exp -\frac{\Delta G_{red}(\phi )}{kT}~\quad~k_{ox}=A\exp -\frac{%
\Delta G_{ox}(\phi )}{kT}.  \label{eqA2}
\end{equation}
For some potential $\phi _{0}$, the two barriers are equal, 
$\Delta G_{red}(\phi _{0})=\Delta G_{ox}(\phi_{0})$, 
and hence $k_{red}=k_{ox}$ such that
the total reaction current $\mathbf{j}$ is zero. When $\phi \neq \phi
_{0}$ (in the figure $\phi > \phi _{0},$ metal is deposited), the barriers
are modified. To first order in $\phi-\phi_0$,
\begin{eqnarray}
\Delta G_{red}(\phi ) &=&\Delta G_{red}(\phi _{0})-\alpha _{red}e(\phi -\phi
_{0}),  \nonumber \\
\Delta G_{ox}(\phi ) &=&\Delta G_{ox}(\phi _{0})+\alpha _{ox}e(\phi -\phi
_{0}),  \label{eqA3}
\end{eqnarray}
with (see Fig.~\ref{FigButlerVolmer} for an explanation), 
\begin{equation}
\alpha _{ox}+\alpha _{red}=1.  \label{eqA4}
\end{equation}
The Butler-Volmer relation then gives the electron transfer current 
\begin{equation}
\mathbf{j}=\mathbf{j}_{0}\left( \exp \left[ \frac{e\alpha _{red}(\phi -\phi
_{0})}{kT}\right] -\exp \left[ -\frac{e\alpha _{ox}(\phi -\phi _{0})}{kT}
\right] \right).  \label{eqA5}
\end{equation}
In our present model, we have supposed in Eq.~(\ref{eqA1}) that the 
reduction of a cation in $\mathbf{k}$ is due to a charge transfer from
the metal site $\mathbf{k+a}$. The link with Eq.~(\ref{eq14}) for the 
reaction rate can then be established. The electrochemical potentials are 
\begin{eqnarray}
\widetilde{\mu }_{\mathbf{k}}^{+}+\widetilde{\mu }_{\mathbf{k+a}}^{e} &=&\mu
_{\mathbf{k}}^{+}+E_{F}-e\Phi _{\mathbf{k+a\rightarrow k}},  \nonumber \\
\widetilde{\mu }_{\mathbf{k}}^{0} &=&\mu _{\mathbf{k}}^{0},  \label{eqA6}
\end{eqnarray}
where we have introduced the potential difference 
\begin{equation}
\Phi _{\mathbf{k+a\rightarrow k}}=V_{\mathbf{k+a}}-V_{\mathbf{k}}-\frac{p_{%
\mathbf{k+a}}^{e}}{e\mathcal{D}(E_{F})}.
\end{equation}
The equilibrium (absence of reaction) is obtained when $\Phi _{\mathbf{%
k+a\rightarrow k}}=\Phi _{\mathbf{k+a\rightarrow k}}^{0}$ for which 
\begin{equation}
\widetilde{\mu }_{\mathbf{k}}^{+}+\widetilde{\mu }_{\mathbf{k+a}}^{e}=\mu _{%
\mathbf{k}}^{+}+E_{F}-e\Phi _{\mathbf{k+a\rightarrow k}}^{0}=\mu _{\mathbf{k}%
}^{0}.
\end{equation}
Furthermore, a reaction rate of the form (\ref{eq14}) can also be written as
\begin{eqnarray}
\sigma _{\mathbf{k,k+a}} &=&w_{\mathbf{k,k+a}}^{\star}\left[ \exp\left({\alpha
_{red}\frac{\widetilde{\mu }_{\mathbf{k}}^{+}+\widetilde{\mu }_{\mathbf{k+a}%
}^{e}-\widetilde{\mu }_{\mathbf{k}}^{0}}{kT}}\right)\right.\nonumber\\
&& \left.\mbox{}-\exp\left({-\alpha _{ox}%
\frac{\widetilde{\mu }_{\mathbf{k}}^{+}+\widetilde{\mu }_{\mathbf{k+a}}^{e}-%
\widetilde{\mu }_{\mathbf{k}}^{0}}{kT}}\right)\right],
\end{eqnarray}
with
\begin{equation}
w_{\mathbf{k,k+a}}^{r}=w_{\mathbf{k,k+a}}^{*}\exp\frac{\alpha _{ox}\left( 
\widetilde{\mu }_{\mathbf{k}}^{+}+\widetilde{\mu }_{\mathbf{k+a}}^{e}\right)
+\alpha _{red}\widetilde{\mu }_{\mathbf{k}}^{0}}{kT}.
\end{equation}
With the help of Eq.~(\ref{eqA6}), we obtain
\begin{eqnarray}
\sigma _{\mathbf{k,k+a}}&=&w_{\mathbf{k,k+a}}^{\star}\left[\exp\left(-{\frac{%
e\alpha _{red}(\Phi _{\mathbf{k+a\rightarrow k}}-\Phi _{\mathbf{%
k+a\rightarrow k}}^{0})}{kT}}\right)\right.  \nonumber \\
& & \quad\left.\mbox{}-\exp\left({\frac{e\alpha _{ox}(\Phi _{\mathbf{%
k+a\rightarrow k}}-\Phi _{\mathbf{k+a\rightarrow k}}^{0})}{kT}}%
\right)\right],
\end{eqnarray}
which has the form of Eq.~(\ref{eqA5}). In the above expressions, for a 
square or a simple cubic lattice, with intersite distance $a$, the
transfer current density is
\begin{eqnarray}
\mathbf{j} & = & a^{d-1}w_{\mathbf{k,k+a}}^{*}\exp\left[{\frac{\alpha
_{ox}\left( \widetilde{\mu }_{\mathbf{k}}^{+}+\widetilde{\mu }_{\mathbf{k+a}%
}^{e}\right) +\alpha _{red}\widetilde{\mu }_{\mathbf{k}}^{0}}{kT}}%
\right]\nonumber\\
& \simeq & a^{d-1}w_{\mathbf{k,k+a}}^{*}e^{\mu_{\mathbf{k}}^{0}/kT},
\end{eqnarray}
where the last expression is valid close to equilibrium. In that case,
and with the help of Eq.~(\ref{eq15}) for $w^{*}_{\mathbf{k,k+a}}$, 
the constant $\mathbf{j}_0$ of the Butler-Volmer law can be identified
as $\mathbf{j}_0=a^{d-1}w^{*}e^{\mu^{0}/kT}$, where $\mu^0$ is the
equilibrium chemical potential for the metal species. Note that with
Eq.~(\ref{eq15}), our expression is only an approximation to the 
Butler-Volmer law, valid close to equilibrium; to get a complete
correspondence with the Butler-Volmer model, a dependence of $w^*$ 
on the electrochemical potentials needs to be introduced.

\end{document}